\documentclass[journal=jctc,manuscript=article,layout=twocolumn]{achemso}

\usepackage{achemso}

\usepackage{graphicx}
\usepackage{dcolumn}
\usepackage{bm}
\usepackage{amsmath}

\usepackage[capitalise]{cleveref}
\SectionNumbersOn
\usepackage{pdfpages}

\title{Machine Learning Diffusion Monte Carlo Energies}

\author{Kevin Ryczko}
 \email{kevin.ryczko@uottawa.ca}
 \affiliation{Good Chemistry Company, Vancouver, British Colombia, Canada, V6E 4B1}

\author{Jaron T. Krogel}
 \affiliation{Materials Science and Technology Division, Oak Ridge National Laboratory, Oak Ridge, Tennessee, United States, 37831}

\author{Isaac Tamblyn}
\email{isaac.tamblyn@uottawa.ca}

\affiliation{Department of Physics, University of Ottawa, Ottawa, Ontario, Canada, K1N 6N5}
  \affiliation{Vector Institute for Artificial Intelligence, Toronto, Ontario, Canada, M5G 1M1}

\date{\today}
\begin{document}

\begin{abstract}
We present two machine learning methodologies that are capable of predicting diffusion Monte Carlo (DMC) energies with small datasets ($\approx$~60 DMC calculations in total). The first uses voxel deep neural networks (VDNNs) to predict DMC energy densities using Kohn-Sham density functional theory (DFT) electron densities as input. The second uses kernel ridge regression (KRR) to predict atomic contributions to the DMC total energy using atomic environment vectors as input (we used atom centred symmetry functions, atomic environment vectors from the ANI models, and smooth overlap of atomic positions). We first compare the methodologies on pristine graphene lattices, where we find the KRR methodology performs best in comparison to gradient boosted decision trees, random forest, gaussian process regression, and multilayer perceptrons. In addition, KRR outperforms VDNNs by an order of magnitude. Afterwards, we study the generalizability of KRR to predict the energy barrier associated with a Stone-Wales defect. Lastly, we move from 2D to 3D materials and use KRR to predict total energies of liquid water. In all cases, we find that the KRR models are more accurate than Kohn-Sham DFT and all mean absolute errors are less than chemical accuracy. 
\end{abstract}

\maketitle


\section{Introduction}
\label{intro}
Determining total internal energies and how they change with respect to some perturbation are central objectives in both quantum chemistry and condensed matter physics. Total internal energies can help identify the stability of an atomistic system, and derivatives  with respect to different variables yield a multitude of desired quantities. Recently, machine learning has been used to calculate total electronic energies for toy models \cite{snyder2012finding, ryczko2018convolutional, ryczko2019deep, ryabov2020neural}, molecular systems \cite{behler2007generalized, rupp2012fast, jager2018machine, schutt2018schnet, brockherde2017bypassing, zhou2019toward}, and solid state systems \cite{faber2015crystal, ryczko2018convolutional, mills2019extensive, ryabov2020neural, ryczko2021orbital}. Thus far, the majority of these works have focused on learning Hartree-Fock (HF) or density functional theory (DFT) computed properties. These models can make accurate and rapid predictions, but their utility is limited to the accuracy of the electronic structure method they were trained upon. When studying atomistic systems, one must properly describe electron correlation to achieve high accuracy. Such a treatment leads to computationally expensive electronic structure methods. For molecules, common techniques include M{\o}ller-Plesset perturbation theory (MP2) or coupled-cluster singles, doubles and perturbative triples (CCSD(T)). For solid-state systems, the gold standard is quantum Monte Carlo (QMC) \cite{foulkes2001quantum}.

Several studies have utilized machine learning to predict quantities computed with MP2 or CCSD(T) \cite{wilkins2019accurate, peyton2020machine, margraf2018making, schran2019automated, bogojeski2020quantum, smith2019outsmarting}. In Ref. \cite{peyton2020machine}, a novel representation, called the density tensor representation, was introduced and was used to predict accurate energies and dipoles of small molecules at the MP2 level. In Ref. \cite{bogojeski2020quantum}, the density $\Delta$-DFT approach was introduced and utilized for small molecules. This methodology used the DFT electronic density as input to a machine learning model to predict a difference in energy between CCSD(T) and DFT, allowing for a molecular dynamics simulation with quantum chemical accuracy. Training on the difference between DFT and CCSD(T), rather than the absolute value, allowed for an improved model accuracy when performing inference (predictions). 

However, machine learning studies for solid-state systems with accurate electronic structure calculations are currently absent in the literature. This is due to the limitations of typical machine learning implementations, the computational cost of the electronic structure calculations, and the lack of abundant QMC data for solid-state systems. Most current machine learning approaches require large quantities of training data, making it infeasible to train on properties calculated with QMC. However, recent machine learning methodologies have eliminated this problem by focusing on scalar functions, rather than scalar quantities \cite{zhou2019toward, ryabov2020neural, ryczko2021orbital}. Namely, with voxel deep neural networks (VDNNs) it was shown that electron kinetic energies could be predicted within chemical accuracy from only 2 DFT calculations using a technique called voxel deep neural networks (VDNNs) \cite{ryczko2021orbital}.

\begin{figure*}
    \centering
    \includegraphics[width=\linewidth]{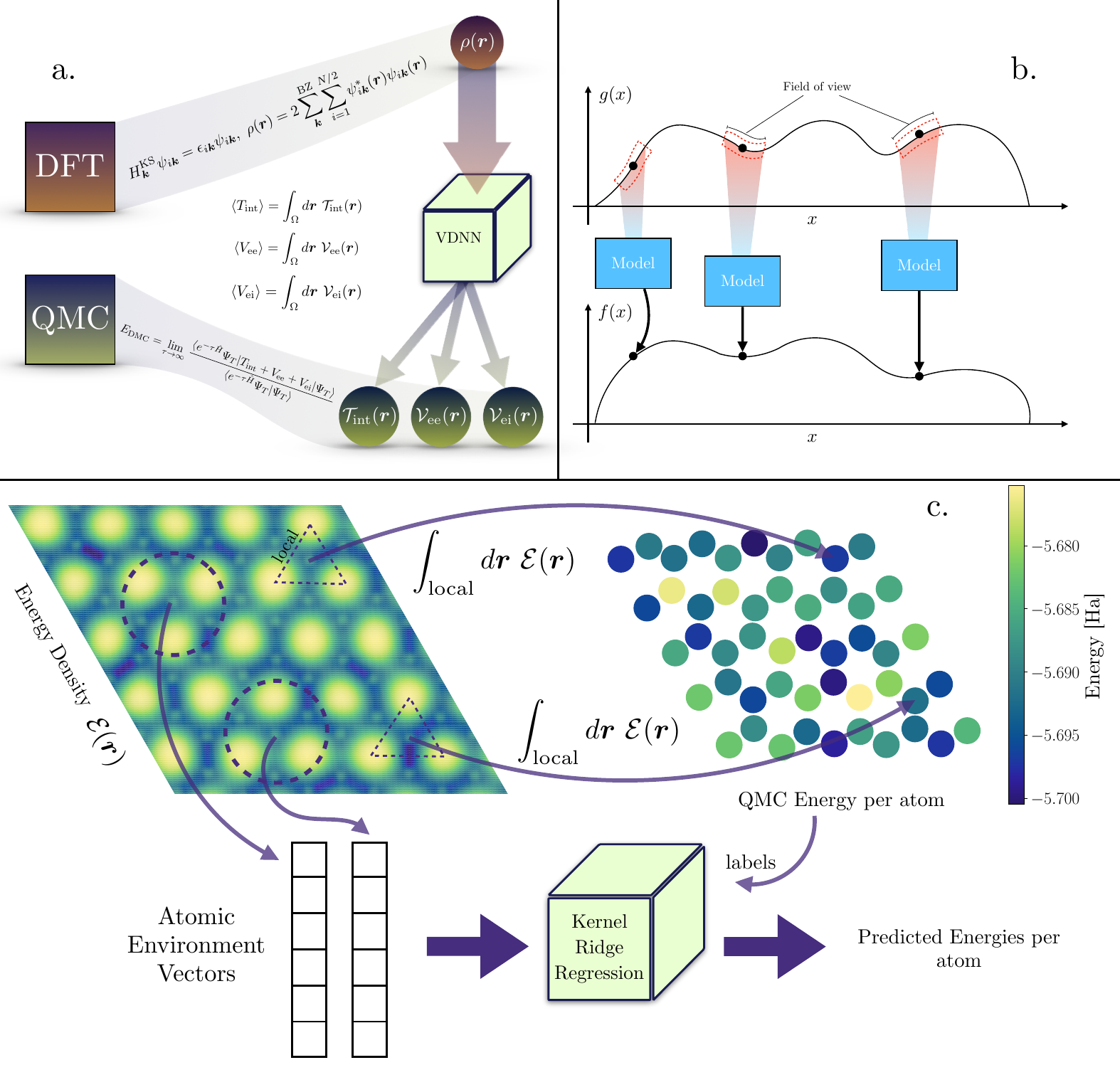}
    \caption{Visual representation of the methodologies used. In (a) we show how a voxel deep neural network is used to map an electron density computed via density functional theory to energy densities computed via diffusion Monte Carlo. These include the interacting kinetic energy density ($\mathcal{T}_{\text{int}}$), the electron-electron energy density ($\mathcal{V}_{\text{ee}}$), and the electron-ion energy density ($\mathcal{V}_{\text{ei}}$). The Hamiltonian $\hat{H}=T_{\text{int}} + V_{\text{ee}} + V_{\text{ei}}$,  $\Psi_T$ is the trial many body wavefunction, and $\tau$ is imaginary time. In (b) we show a 1D example of a voxel deep neural network. A model is used to map the function $g(x)$ to another function $f(x)$. For every value $x$, there is a fixed field of view that is used to construct inputs for the machine learning model. In (c) we show a visualization of how atomic contributions to the total energy are computed and a schematic of  the kernel ridge regression (KRR) training process.}
    \label{fig:fig_1}
\end{figure*}

In this report, we present and compare two methodologies to predict diffusion Monte Carlo (DMC) total energies using small datasets. The first methodology uses VDNNs to compute QMC energy densities by mapping an electron density computed via DFT to kinetic, electron-electron, and electron-ion energy densities computed with diffusion Monte Carlo (DMC). The second approach uses kernel ridge regression (KRR) with atomic environment vectors as inputs to predict atomic contributions to the DMC total energy. We first compare these methodologies for pristine graphene lattices. Afterwards, we study the generalizability of the KRR method, and predict the energy barrier associated with a Stone-Wales defect in a graphene lattice. Lastly, we move from 2D to 3D materials and use a KRR model to predict DMC total energies of liquid water along a molecular dynamics trajectory. To date, this is the first report that calculates QMC energies for solid-state systems with supervised machine learning.

\section{Methods}
\label{methods}
In Kohn-Sham DFT \cite{kohn1965self} the electron density is written as
\begin{equation}
\label{rho}
    \rho(\bm{r}) = 2\sum_{n}^{\text{occ}}\sum_{\bm{k}}w_{\bm{k}} \psi_{n, \bm{k}}^*(\bm{r})\psi_{n, \bm{k}}(\bm{r}).
\end{equation}
In Equation \ref{rho}, $n$ is the band index, $\bm{k}$ is the k-point, $w_{\bm{k}}$ is the weighting associated with the k-point, and $\psi$ is a Kohn-Sham orbital. For the graphene lattices we used a non-orthorhombic supercell with lattice vectors $a_1 = (12.310, 0, 0)$, $a_2 = (-6.155, 10.661, 0)$, and $a_3=(0, 0, 6.000)$ \AA. For liquid water, we took 32 equally spaced snapshots from a previously used dataset \cite{PhysRevLett.120.036002} that consisted of 1000 atomic configurations from 200 ps of path integral molecular dynamics with 32 water molecules. The 200 ps were run under constant atmospheric pressure and were subdivided amongst 6 replicated runs, each with a different temperature (300 - 350 K). We refer the reader to the original manuscript for more information \cite{PhysRevLett.120.036002}. The QMC electronic contribution to the energy density, as derived in Ref. \cite{krogel2013quantum} is 
\begin{equation}
    \label{qmc}
    \mathcal{E}(\bm{r}) = \mathcal{T}(\bm{r}) + \mathcal{V}_{\text{ee}}(\bm{r}) + \mathcal{V}_{\text{ei}}(\bm{r}),
\end{equation}
where the many-body kinetic energy density is
\begin{equation}
\label{mbke}
    \mathcal{T}(\bm{r})=-\frac{1}{2}\sum_{i=1}^N\int d\bm{R}~\delta(\bm{r} - \bm{r}_i)\Psi^*(\bm{R})\nabla_i^2\Psi(\bm{R}),
\end{equation}
and the many-body potential energy densities are
\begin{equation}
    \label{mbpot}
    \mathcal{V}_\text{ab}(\bm{r})=\frac{1}{2}\sum_{i=1}^{N_a}\sum_{j=1}^{N_b}\int d\bm{R}~ \delta(\bm{r} - \bm{r}_i)|\Psi(\bm{R})|^2v_{\text{ab}}(\bm{r}_i, \bm{r}_j).
\end{equation}
In Equations \ref{mbke} and \ref{mbpot}, $\bm{R}\equiv\{\bm{r}_1, \bm{r}_2, \hdots, \bm{r}_N\}$ is the full many-body space, $v_{\text{ab}}$ is the interaction potential between two species, and $\Psi$ is the many-body wavefunction. For $\mathcal{V}_\text{ee}$, the interaction potential $v_\text{ee}$ is the Coulomb potential. For the potential energy density term $\mathcal{V}_\text{ei}$, the interaction potential is a norm-conserving pseudopotential of the Burkatzki-Filippi-Dolg form \cite{burkatzki2007energy}. The many-body wavefunction is constructed from a Slater determinant of Kohn-Sham orbitals and an optimized Jastrow factor. The QMC energy densities are generated in 3 stages. The first stage is a DFT calculation to obtain the Kohn-Sham orbitals needed for the Slater determinant. To perform the DFT calculation, we used Quantum Espresso \cite{giannozzi2009quantum} included with QMCPACK \cite{kim2018qmcpack} with an energy cut-off of 150 Ha. For the graphene lattices we used a $4\times4\times1$ k-point grid and for liquid water we used a $2\times2\times2$ k-point grid. The second stage is the Jastrow factor optimization. We used a one-body Jastrow factor for the nuclei and a two-body Jastrow factor for the electrons. The final stage is evaluating the energy which was done in 3 steps. This included a variational Monte Carlo (VMC) calculation followed by two DMC calculations, each with different time steps. We refer the reader to the supplemental information (SI) for more information on the Jastrow optimization and other QMC parameters \cite{si}. All three stages were performed at each k-point, and the final energy densities were obtained by averaging over all k-points. All calculations were performed using QMCPACK \cite{kim2018qmcpack}, Nexus \cite{krogel2016nexus}, and a custom Python library \cite{github-qmc}.  The atomic energies are computed from the energy density which is stored as a 3D array. The distance between a particular voxel and each atom is computed using the minimum image convention. The voxel is then assigned to the closest atom. The assigned voxels are then integrated to obtain the energy per atom. This yielded 50 atomic contributions per calculation for graphene and 96 atomic contributions per calculation for liquid water. By using atomic energy contributions, one has access to $M\times\langle N\rangle$ training points, where $M$ is the number of calculations, and $\langle N \rangle$ is the average number of atoms per supercell. If using total energies alone, one only has access to $M$ values for training.

When generating a dataset for VDNNs, we used $\approx 5\times10^{5}$, $19\times19\times19$ slices of the DFT electron densities that were collected from the training configurations such that a uniform distribution of the function $\mathcal{F} = \sqrt{\mathcal{T}^2 + \mathcal{V}_{\text{ee}}^2 + \mathcal{V}_{\text{ei}}^2}$ was produced. These parameter choices were found to be optimal in a previous study on the prediction of DFT kinetic energy densities \cite{ryczko2021orbital}. Of these electron density slices, 99\% were used for training and 1\% were used for validation. When training the VDNNs, we map the electron density slices to values of $\mathcal{T},~\mathcal{V}_{\text{ee}}$, and $\mathcal{V}_{\text{ei}}$, as shown in \cref{fig:fig_1}a. We used the convolutional neural network (CNN) from Ref. \cite{ryczko2021orbital}, a batch size of 512, a learning rate of $10^{-5}$, the Adam optimizer \cite{kingma2014adam}, and layer-wise adaptive rate scaling with clipping \cite{you2017scaling}. We use ensemble models, where 5 independent networks were trained simultaneously and total energies were averaged over the models. Similar to Ref. \cite{ryczko2021orbital}, a rigid shift was applied to the predictions based on the residuals of the training set. This shift is needed to compensate for the errors from the model that accumulate when integrating on large grids. The shift can be calculated by computing the average error within an energy window. The number of energy windows depends on the structure of the residuals, and therefore differs for each model within the ensemble. We used a grid search to compute the optimal number of energy windows, and found the number of energy windows to be in the range $[8,32]$. We trained the networks for 250 epochs, and we found that the models began to overfit beyond $\approx250$ epochs. This overfitting can be attributed to the stochastic nature of the underlying training set. Beyond a certain point in training, the model begins to fit to statistical noise. We stopped training prior to this and used the model with the lowest mean squared error on the validation set. Training and inference was done across multiple nodes, each with 4 NVIDIA V100 GPUs. Our codes can be found here \cite{github-qmc}.

For learning the atomic energy contributions, we tried various atomic environment descriptors including atom centred symmetry functions (ACSF) \cite{behler2011atom}, atomic environment vectors (AEV) from the ANI models \cite{smith2018less, smith2019outsmarting, devereux2020extending},  and smooth overlap of atomic positions (SOAP) \cite{bartok2013representing}. We have listed the parameters for these descriptors in the SI \cite{si}. Apart from AEV, we used the Python library dscribe \cite{dscribe} for generating the atomic environment descriptors with the default parameters. For AEV, we used the parameters which can be found in torchani \cite{gao2020torchani}. For the machine learning model, we employed Gradient Boosted Decsion Trees (GBDT), Gaussian Process Regression (GPR), artificial neural networks (ANNs), and KRR. We used the Python library scikit-learn \cite{scikit-learn} to implement all models and tried various sets of parameters for each model using grid searches. For both the graphene lattices and liquid water, we found that the KRR models had the lowest mean absolute errors (MAEs). These results can be found in the SI \cite{si}.

\begin{figure}[ht]
    \centering
    \includegraphics[width=\linewidth]{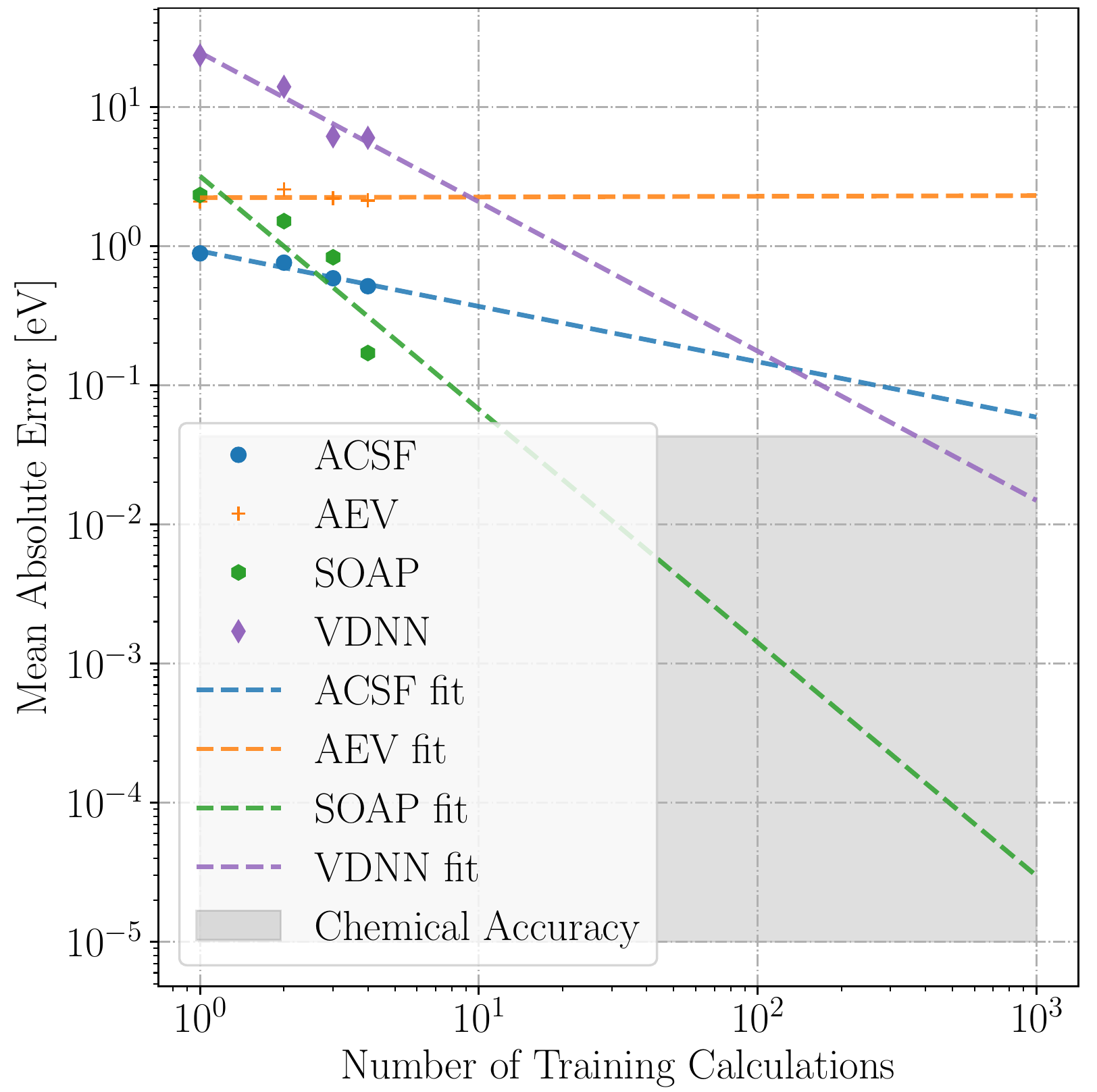}
    \caption{Learning curves for various models along with their respective linear fits. The linear fits allows one to estimate how much data is necessary to achieve a particular accuracy. We estimate for VDNNs that one would need $\approx 400$ calculations to achieve chemical accuracy. For KRR with SOAP as the atomic environment vectors, one would only need $\approx 12$ calculations. However, in practice one may find saturation of the model error as the training set size increases.}
    \label{results_1}
\end{figure}

\begin{figure}[ht]
    \centering
    \includegraphics[width=\linewidth]{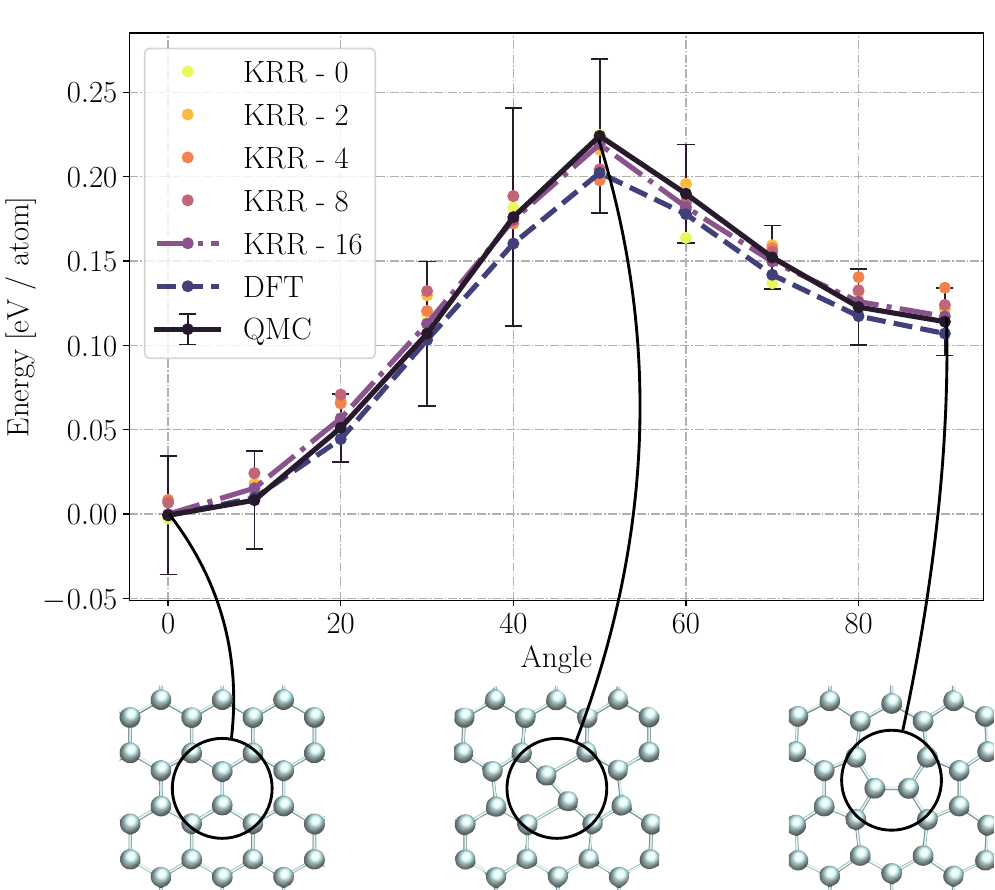}
    \caption{The energy barrier curve associated with creating a Stone-Wales defect in a pristine graphene lattice. The error bars for QMC show the standard deviation of the calculation. The model KRR-16 has the lowest mean absolute error with 4.17 meV / atom.}
    \label{results_2}
\end{figure}

\section{Results}
\label{results}

\subsection{Comparing Voxel Deep Neural Networks and Kernel Ridge Regression}
The first task is the prediction of DMC total energy for pristine graphene lattices with random atomic perturbations. When building this dataset, we generated 8 atomic configurations where random perturbations of the atomic positions were applied to the equilibrium structure. To generate the random displacements, uniform random numbers were drawn with the range $[-0.1, 0.1]$ and were added to the pristine coordinates. 4 of these calculations were used to generate training/validation data and the remaining 4 were used for testing. For the VDNNs, we trained 4 independent ensemble models, where each model was allowed to see 1, 2, 3 or 4 training configurations. We then computed the total energies of the testing set by integrating the predicted energy densities from the ensemble models. In \cref{results_1}, we plot the learning curve for the VDNNs. The best MAE found for VDNNs was 120 meV / atom. For both KRR and VDNNs, the learning curve allows one to estimate the MAE of a model if additional calculations were available. A  linear fit was done in order to obtain these estimates. In \cref{results_1}, these estimates are labelled by the keyword ``fit.'' If one would like to have a VDNN that is capable of making predictions within chemical accuracy, one would need $\approx~400$ additional calculations. For the KRR models, we found that the SOAP representation had the lowest MAE which was 3.40 meV / atom. In comparison to the VDNNs, this is a 35x improvement. In addition, we estimate that we would need $\approx~8$ additional calculations to obtain a model capable of making DMC energy predictions within chemical accuracy. This accuracy, however, is limited to pristine graphene lattices with atomic perturbations and model saturation may occur. To understand which method is more transferable, we chose to make predictions for a unit cell of diamond (orthorhombic unit cell with 8 atoms). The electron density was also computed using Quantum Espresso \cite{giannozzi2009quantum} using an energy cut-off of 150 Ha, a $12\times12\times12$ k-point grid, and a lattice constant of 3.57 \AA. Since KRR does not include a standard deviation, we chose to use GPR. For VDNNs, we considered the standard deviation of the ensemble prediction. We found that the standard deviation from GPR (2039.7 Ha) was two orders of magnitude larger than the standard deviation from VDNNs (25.9 Ha). Despite both standard deviations being quite large, this result indicates that VDNNs are more transferable. However, the standard deviation from the VDNNs is based on the ensemble and there could be cases where the error is large but the standard deviation is small (or vice versa). One should be cautious when extrapolating. In addition, one could also train directly on the total energy rather than the atomic energy contributions to the total energy. We used various models and atomic descriptors (concatenated for all atoms yielding 1 input vector per structure) and we found that learning the energy contributions yielded errors lower by an order of magnitude. For more information, we refer the reader to the SI \cite{si}.

\begin{figure*}[ht]
    \centering
    \includegraphics[width=\linewidth]{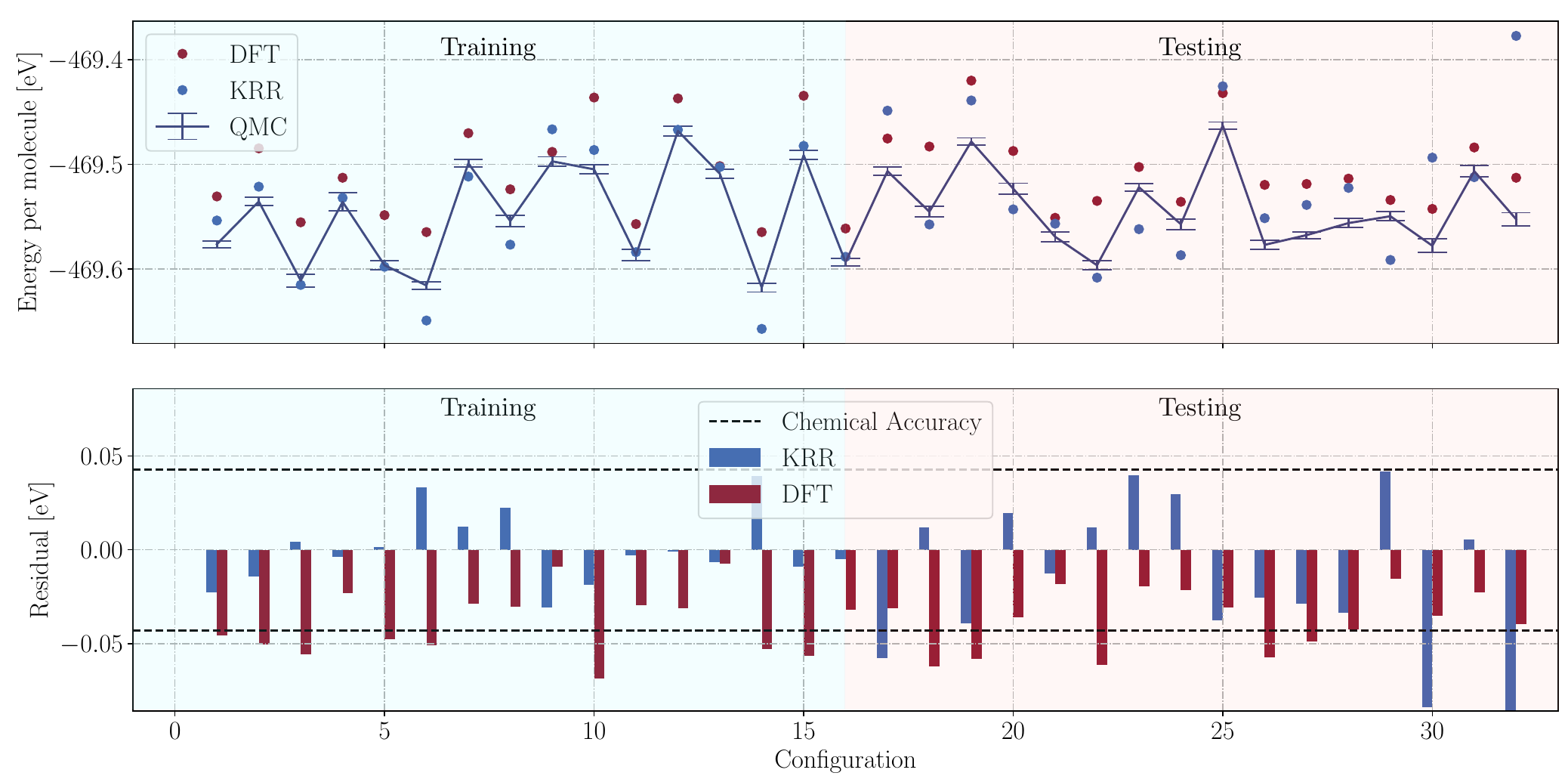}
    \caption{Top: QMC, DFT, and KRR model prediction energies for liquid water. Bottom: Residuals for DFT and the KRR model. The KRR model shows a 1.38x improvement over DFT with a MAE of 38.1 meV / molecule.}
    \label{results_3}
\end{figure*}

\subsection{Predicting the Stone-Wales Energy Barrier}
Given the performance of KRR in comparison to VDNNs, we concentrate on analyzing the capabilities of KRR in more complex scenarios. Namely, we are interested in studying the generalizability of the KRR model in regards to calculating the energy barrier curve associated with the introduction of a Stone-Wales defect within a graphene lattice, which was previously investigated with DFT in Ref. \cite{li2005defect}. As shown in \cref{results_2}, by rotating two atoms in the graphene lattice, one transitions from a pristine lattice ($\theta=0^{\circ}$) to a lattice with a Stone-Wales defect ($\theta=90^{\circ}$). For the training set, we started with the 3 configurations shown in \cref{results_2}. These configurations do not have any random atomic perturbations added. We then ran 16 additional DMC calculations of graphene lattice with a single Stone-Wales defect and atomic perturbations. The testing dataset consisted of 10 configurations, where two atoms were rotated such that the $nth$ configuration had an angle $\theta = n \pi/ 18,~n\in[1,9]$. For each configuration, a DFT structural relaxation was performed while holding the two rotated atoms fixed. As done in Ref. \cite{li2005defect}, the bond length between the two rotated atoms was also reduced. We first trained a KRR model ignoring the 16 additional DMC calculations and predicted the energy barrier curve. In \cref{results_2}, this is labelled by ``KRR-0.'' With only 3 total energy reference points, the KRR model does exceptionally well, with a MAE of 7.32 meV / atom. We then added in 2, 4, 8, or 16 additional DMC calculations (labelled ``KRR-2'', ``KRR-4'', ``KRR-8'', and ``KRR-16'' respectively in \cref{results_2}), trained a KRR model, and made predictions for the energy barrier. Interestingly, when comparing the models KRR-0 and KRR-[2,4,8], the MAEs increase (up to 1.8x worse with KRR-4) with the addition of data. We argue that with the addition of new data, there are two areas of chemical space that must be accurately modelled and connected in a physically meaningful way. The models KRR-[2,4,8] do not connect these two areas of chemical space well, leading to higher MAEs. However, with all 16 additional points, the MAE becomes 4.23 meV / atom (1.7x times improvement from KRR-0). The KRR-16 model learned to generalize its knowledge from the additional calculations to make accurate predictions of the energy barrier.

 We define the difference in energy between the highest peak seen in \cref{results_2} and the energy at $\theta = 0$ to be the energy barrier needed to be overcome to transition from a pristine lattice to one with a Stone-Wales defect. We find the energy barrier to be 224(46) meV / atom based on the DMC calculation. DFT underestimates the energy barrier by 17.6 meV / atom and KRR-16 underestimates the energy barrier by only 5 meV / atom.

 \subsection{Predicting Energies of Liquid Water}

 We now move from 2D to 3D materials and study the performance of KRR when applied to liquid water. Since there are two atom types, one KRR model was trained to predict atomic energy contributions for H, and another model was trained to predict atomic energy contributions for O. In both cases we found that the AEV representation had the lowest MAE in comparison to other representations. As described in \cref{methods}, there were a total of 32 DMC calculations that were done. Half of these were used in training, and the remaining half were used as a test set. In \cref{results_3}, these data points are highlighted under ``Training'' (in cyan) and ``Testing'' (in light red). For KRR, most of the predictions are within chemical accuracy. The MAE of the test set was 27.6 meV / molecule. In comparison to DFT, which had a MAE of 38.1 meV / molecule, the KRR model shows a 1.38x improvement.
 
 \subsection{Limitations and Proposed Future Work}
Before concluding, we discuss the limitations and future work of both methods presented here. Firstly, both VDNNs and KRR models are limited to the data they are trained with (as is the case with all machine learning models). 

\subsubsection{Voxel Deep Neural Networks}
For machine learning models that predict scalar quantities (i.e. total energy), there is no error correction that must be done. However, as discussed previously, error from VDNNs accumulates when integrating, and must be accounted for. Secondly, with VDNNs there is no dependence between neighbouring pixels that are output by the model. The dependence could be applied by outputting neighbouring pixels and averaging over the predictions or in some sort of post-processing routine. Thirdly, due to use of images, rotational invariance is not inherently built-in to the model, but must be learned during training. When constructing diversity in a dataset for VDNNs, one must consider the structure of the electron density, rather than a chemical environment. The two are inter-linked, but one must consider how the electron density images change within a dataset, and use this information when applying it to a new system.  Lastly, since VDNNs must make millions of inference calls for a 3D grid, prediction time is relatively slow (on the order of minutes) compared to machine learning methods that output a scalar quantity. Despite these limitations, a promising example of machine learning energy densities with VDNNs is predicting an exchange-correlation energy correction to a DFT calculation. Future work of ours involves training a model on energy density differences between DMC and DFT. By obtaining the difference between the interacting and non-interacting kinetic energy as well as the difference between the true electron interaction energy and the Hartree energy one has access to the true exchange-correlation energy. 

\subsubsection{Kernel Ridge Regression}
The advantage of predicting atomic contributions to the total energy is rapid inference. The KRR models take seconds to train and perform inference, and have rotational and translational symmetry built in. However, our current KRR implementation does not have automatic differentiation. Automatic differentiation with kernel methods has been done previously \cite{gardner2018gpytorch}, and future work of ours involves using these packages to obtain forces at the DMC level. In addition, this work shows that it is possible to perform supervised learning with DMC data. Given that our KRR method trains on atomic contributions to the total energy, we expect that size of datasets needed for training DNNs will be smaller by a factor of $\langle N \rangle$, which is the average number of atoms per calculation. Other future work of ours involves $\Delta$-learning (from DFT to DMC) atomic contributions to the total energy.

\section{Conclusion}
We have used two machine learning methodologies to predict DMC total energies. The first methodology we used was VDNNs to map electron densities computed with DFT to energy densities calculated with DMC for pristine graphene lattices. After comparing these results with our second methodology, which used KRR and atomic environment vectors to predict atomic contributions to the total energy, we found that the errors from KRR were an order of magnitude lower. Given this result we then further studied the capabilities of KRR to predict atomic contributions to the energy for more complex tasks. This included predicted the energy barrier when introducing a Stone-Wales defect into a pristine graphene lattice and studying liquid water. In both cases, we found that KRR trained on DMC outperforms DFT. In addition, the KRR method requires no initial DFT calculation, but only the computation of atomic environment vectors. This allows one to rapidly infer energies at the DMC level. Since our method trains on atomic contributions to the total energy, we also conclude that dataset sizes in future studies can be reduced by a factor of $\langle N \rangle$, which is the average number of atoms per calculation. This would allow for DNNs to be used to predict DMC energies, furthering the performance showed in this report. Our present results and the continuous increase of compute power allows for the increase of the production of QMC datasets. The increase of QMC data allows for more sophisticated models enabling the use of models trained on accurate electronic structure methods in a practical setting. However, our present models are at the ``proof-of-concept" stage and are step towards having machine learning models trained on accurate electronic structure calculations. An addition, further analysis of the models would have to be done before using them in practical applications.

\section{Acknowledgements}
The Authors acknowledge Compute Canada, the Vector Institute for Artificial Intelligence, and the National Energy Research Scientific Computing Center for computational resources. KR and IT acknowledges the National Sciences and Engineering Council of Canada for funding. Work performed by JK (mentorship) was supported by the US Department of Energy, Office of Science, Basic Energy Sciences, Materials Sciences and Engineering Division.

\section{Supplemental Information}
In the supplemental information, we include additional details about the QMC calculations, the parameters used along with the Python code to define the atomic environment descriptors, as well as a table of results comparing different approaches, models, and descriptors for the systems reported in the manuscript.
\bibliography{refs}

\begin{figure}[ht]
    \centering
    \includegraphics[width=\linewidth]{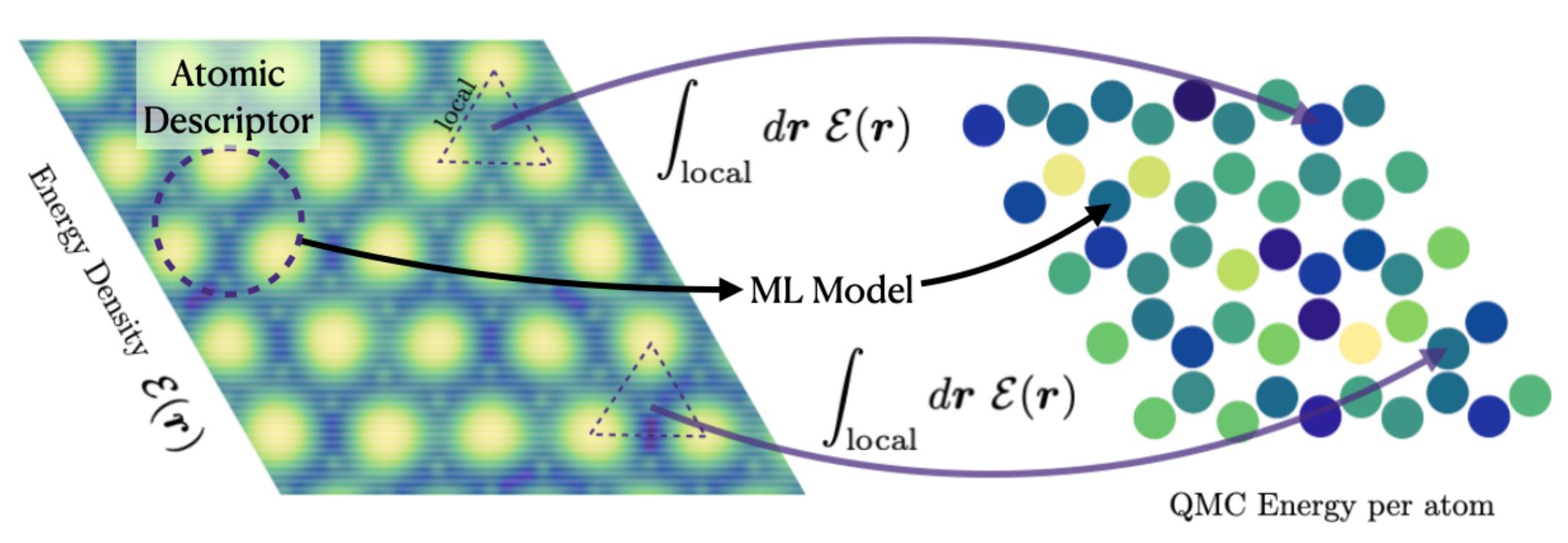}
    \caption{TOC Graphic}
\end{figure}



\section*{Supplementary material}

# Machine Learning Diffusion Monte Carlo Energies - Supplemental Information

Kevin Ryczko,*,† Jaron T. Krogel,‡ and Isaac Tamblyn*,¶

†Good Chemistry Company, Vancouver, British Colombia, Canada, V6E 4B1

‡Materials Science and Technology Division, Oak Ridge National Laboratory, Oak Ridge, Tennessee, United States, 37831

¶Department of Physics, University of Ottawa, Ottawa, Ontario, Canada, K1N 6N5

§Vector Institute for Artificial Intelligence, Toronto, Ontario, Canada, M5G 1M1

E-mail: kevin.ryczko@uottawa.ca; isaac.tamblyn@uottawa.ca

## More Information on the QMC Calculations

### Jastrow Factors

The Jastrow factors were represented by B-splines and the coefficients were optimized using the quartic optimizer in QMCPACK.[1] The meshes used in the QMC calculations were increased by 20% and we found the average and corresponding standard deviation of the variance to energy ratio of $0.0205 \pm 0.0014$. The optimization was run for 100 blocks each consisting of 51200 Monte-Carlo samples.

### Calculation of the Energy Densities

In this subsection, we provide additional details for the calculation of the energy densities. The first step is performing variational Monte Carlo (VMC) for 40 blocks with 30 warm-up

1

steps and 10 steps used for averaging. Warm-up steps are steps taken where data is excluded from calculating quantities. The second step was DMC with a larger time step (0.02) for 20 blocks with 20 warm-up steps and 5 steps used for averaging. The final step was DMC with a smaller time step (0.01) for 200 blocks with 20 warm-up steps followed by 10 steps. The total number of Monte-Carlo samples completed for each of the final DMC calculations was 20 blocks $\times$ 10 steps $\times$ 1024 walkers $\times$ 64 MPI processes $\approx 13 \times 10^6$. The average and corresponding standard deviation of the total energy error across all calculations was $0.426 \pm 0.083$ meV.

## Parameters Used for Atomic Environment Descriptors

We employed various atomic environment descriptors including atom centred symmetry functions (ACSF),[2] atomic environment vectors (AEV) from the ANI models,[3–5] and smooth overlap of atomic positions (SOAP).[6] Apart from AEV, we used the Python library dscribe[7] for generating the atomic environment descriptors with the default parameters. For AEV, we used the parameters which can be found in torchani.[8] For clarity, we show the respective Python codeblocks for generating each descriptor. For ACSF, we used:

```python
from dscribe.descriptors import ACSF
cm = ACSF(
    species=["C"],
    rcut=6.0,
    g2_params=[[1, 1], [1, 2], [1, 3]],
    g4_params=[[1, 1, 1], [1, 2, 1], [1, 1, -1], [1, 2, -1]],
    periodic=True)
```

For AEV, we used:

```python
import torch
```

```
import torchani
device='cpu'
Rcr = 5.2000e+00
Rca = 3.5000e+00
EtaR = torch.tensor([1.6000000e+01], device=device)
ShfR = torch.tensor([9.0000000e-01, 1.1687500e+00, 1.4375000e+00,
1.7062500e+00, 1.9750000e+00, 2.2437500e+00, 2.5125000e+00,
2.7812500e+00, 3.0500000e+00, 3.3187500e+00, 3.5875000e+00,
3.8562500e+00, 4.1250000e+00, 4.3937500e+00, 4.6625000e+00,
4.9312500e+00], device=device)
Zeta = torch.tensor([3.2000000e+01], device=device)
ShfZ = torch.tensor([1.9634954e-01, 5.8904862e-01,
9.8174770e-01, 1.3744468e+00, 1.7671459e+00, 2.1598449e+00,
2.5525440e+00, 2.9452431e+00], device=device)
EtaA = torch.tensor([8.0000000e+00], device=device)
ShfA = torch.tensor([9.0000000e-01, 1.5500000e+00,
2.2000000e+00, 2.8500000e+00],
    device=device)
species_order = ['H', 'O']
num_species = len(species_order)
aev_computer = torchani.AEVComputer(Rcr, Rca, EtaR, ShfR,
    EtaA, Zeta, ShfA, ShfZ, num_species)
```

For SOAP, we used:

```
from dscribe.descriptors import SOAP
cm = SOAP(species=['C'], periodic=True, rcut=6., nmax=8, lmax=6)
```

3

## Comparing Different Machine Learning Methods and Atomic Representations

See Table 1.

Table 1: Mean absolute Errors (MAE) in meV / atom for various condensed systems, machine learning models, atomic descriptors, and methodologies. Refer to the main text for the acronyms. The descriptions $\sum_i E_i$ signifies total energy predictions found performing a summation over the energy contributions (from the energy density) and $E_{total}$ signifies predicting the total energy directly.

| System | Model | Atomic Descriptor | Description | MAE [meV / atom] |
|--------|-------|-------------------|-------------|------------------|
| Graphene | KRR | ACSF | $\sum_i E_i$ | 10.28 |
| Graphene | KRR | AEV | $\sum_i E_i$ | 42.47 |
| Graphene | KRR | SOAP | $\sum_i E_i$ | **3.40** |
| Graphene | GBDT | ACSF | $\sum_i E_i$ | 19.06 |
| Graphene | GBDT | AEV | $\sum_i E_i$ | 42.47 |
| Graphene | GBDT | SOAP | $\sum_i E_i$ | 18.58 |
| Graphene | GPR | ACSF | $\sum_i E_i$ | 45.13 |
| Graphene | GPR | AEV | $\sum_i E_i$ | 43.39 |
| Graphene | GPR | SOAP | $\sum_i E_i$ | 11.49 |
| Graphene | ANN | ACSF | $\sum_i E_i$ | 580.4 |
| Graphene | ANN | AEV | $\sum_i E_i$ | 1058.2 |
| Graphene | ANN | SOAP | $\sum_i E_i$ | 4361.8 |
| Graphene | KRR | ACSF | $E_{total}$ | 42.34 |
| Graphene | KRR | AEV | $E_{total}$ | 42.52 |
| Graphene | KRR | SOAP | $E_{total}$ | 41.98 |
| Stone-Wales | KRR | SOAP | $\sum_i E_i$ | **4.17** |
| Stone-Wales | GBDT | SOAP | $\sum_i E_i$ | 23.56 |
| Stone-Wales | GPR | SOAP | $\sum_i E_i$ | 13.95 |
| Stone-Wales | ANN | SOAP | $\sum_i E_i$ | 1000.9 |
| Water | KRR | ACSF | $\sum_i E_i$ | 16.39 |
| Water | KRR | AEV | $\sum_i E_i$ | **6.89** |
| Water | KRR | SOAP | $\sum_i E_i$ | 13.79 |
| Water | GBDT | ACSF | $\sum_i E_i$ | 14.78 |
| Water | GBDT | AEV | $\sum_i E_i$ | 15.52 |
| Water | GBDT | SOAP | $\sum_i E_i$ | 16.83 |
| Water | GPR | ACSF | $\sum_i E_i$ | 16.17 |
| Water | GPR | AEV | $\sum_i E_i$ | 11.60 |
| Water | GPR | SOAP | $\sum_i E_i$ | 11.02 |
| Water | ANN | ACSF | $\sum_i E_i$ | 582.1 |
| Water | ANN | AEV | $\sum_i E_i$ | 2663.8 |
| Water | ANN | SOAP | $\sum_i E_i$ | 2313.0 |
| Water | KRR | ACSF | $E_{total}$ | 17.75 |
| Water | KRR | AEV | $E_{total}$ | 43.80 |
| Water | KRR | SOAP | $E_{total}$ | 14.90 |

6



\end{document}